\documentclass{elsart1p}
\usepackage{graphicx}
\usepackage{amssymb}
\begin{document}
\begin{frontmatter}
%
%
\title{Measurement of Transverse Single Spin Asymmetry A$_N$ in Eta Mass Region
at Large Feynman X$_F$ 
with the STAR Forward Pion Detector}
%
%
%
\author{ Steven Heppelmann, for the STAR Collaboration }
\address{Davey Lab,Penn State University, University Park, Pa. 16802}
\begin{abstract}
The large values of the transverse single spin asymmetry, $A_N$, 
seen in forward $\pi^0$ production from polarized proton collisions 
have stimulated important questions and have been studied in QCD 
based transverse spin models.  We report the first RHIC  measurement of $A_N$ 
for the two photon mass region of the $\eta$ meson. 
The $\eta$ peak for meson energies greater than 50 GeV 
(Feynman $x_F >0.5$)  was observed in the STAR Forward Pion Detectors, 
along with 
$\pi^0$ mesons, at pseudo-rapidity  of 3.65 in $\sqrt{s}=200 GeV$ pp
 collisions. The $\pi^0$ transverse asymmetry, which has already been 
reported in detail by the STAR collaboration, is compared to the
transverse asymmetry in the $\eta$ mass region. 
The surprising observation is that in this kinematic 
region, the $\eta$ asymmetry is larger than the already large
 $\pi^0$ asymmetry. 

\end{abstract}
\end{frontmatter}
%
\section{Introduction}

Measurements of large transverse single spin asymmetries of high energy 
forward meson production in collisions between polarized protons have 
surprised physicists for many years. It is well known that such asymmetries 
require both a non-vanishing helicity flip amplitude and 
interference between real and imaginary amplitudes. From a perturbative 
QCD perspective, this means that these asymmetries do not occur in
leading twist QCD with collinear factorization. 
It has been determined, however, that the $\pi^0$ spin
averaged cross section is in fairly good agreement with Next to Leading Order
Perturbative QCD\cite{Adams:2006uz}. Thus these transverse asymmetries provide 
a laboratory for the sensitive study of non-leading twist aspects of 
hadron interactions in processes where the underlying leading twist 
calculations have been shown to be meaningful. 

Two basic parton model pictures have developed to explain how these 
asymmetries can come about. 
In the Sivers model\cite{Sivers:1990fh,D'Alesio:2004up}, 
the initial transverse  momentum distribution of large x quarks, 
in conjunction with absorptive 
effects, could lead to the required spin dependent transverse momentum that 
can explain these  asymmetries. 
In the Collins model\cite{Collins:1993kq}, corresponding 
higher twist effects are 
associated with the final state fragmentation process that produces the final 
state meson from a parton. The parton retains its initial state 
polarization through the hard scattering process.

Here we will consider a comparison between the forward asymmetries, $A_N$,  
measured with forward $\pi^0$ and $\eta$ mesons. 
The central assumption of factorization, implicit in either
the Collins or Sivers approach, leads to the conclusion
that the hard scattering process
determines the transverse momentum of the jet, independent of 
fragmentation details.
However, jets of different types may favor fragmentation into 
different mesons. For example, it is widely accepted that if a u quark
fragments with large z, it tends to 
fragment into a $\pi^+$ meson and a d quarks similarly tends to fragment into 
a $\pi^-$ meson. It is believed that the transverse spin of the 
quarks (like the longitudinal spin) in a polarized proton is highly 
correlated with isospin of those quarks. The well known difference in sign of
$A_N$ for $\pi^+$ vs. $\pi^-$ mesons may be evidence of this. 
When we flip the z 
component of the isospin ($I_z$) of the observed meson, we also tend to
flip the transverse 
asymmetry $A_N$ and apparently the isospin of the jet parton. 
If the pion $I_z$
is highly correlated with $A_N$, it is natural to ask how, for the $I_z=0$
mesons, does the 
asymmetry depend upon the total isospin $I$. Does the transverse 
single spin asymmetry differ for the $I=0$ $\eta$ meson and
the $I=1$ $\pi^0$ meson?
Of course the $\eta$ and $\pi^0$ differ not only in isospin but also
in the strangeness content of the wave function,
\begin{eqnarray}
\pi^0=\frac{1}{\sqrt{2}}(u\bar{u}-d\bar{d}) & \hspace{1.in}&
\eta \simeq \frac{1}{\sqrt{3}}(u\bar{u}+d\bar{d}-s\bar{s})
\end{eqnarray}
where to obtain the simple expression for $\eta$, 
the $\eta,\eta'$ mixing angle has been approximated 
as $\theta_P \simeq -19.4^\circ$ \cite{Bramon:1997va}. 
While QCD based models for transverse single spin asymmetries have
been widely studied, these models have not yet been extended to make 
predictions for differences in $A_N$ for $\eta$ vs. $\pi^0$ production.
 
Observed differences in $A_N$ for $\pi^0$ and $\eta$ pseudo-scalar mesons 
have already been reported by E704 \cite{Adams:1997dp}. 
For Feynman $x_F$ above 0.4, they 
reported a substantially larger asymmetry for $\eta$ than $\pi^0$ mesons 
but the significance of the difference was marginal.

\section{Measurement}

\begin{figure}[ht]
\begin{minipage}[b]{3.0in}
\includegraphics*[width=2.9in]{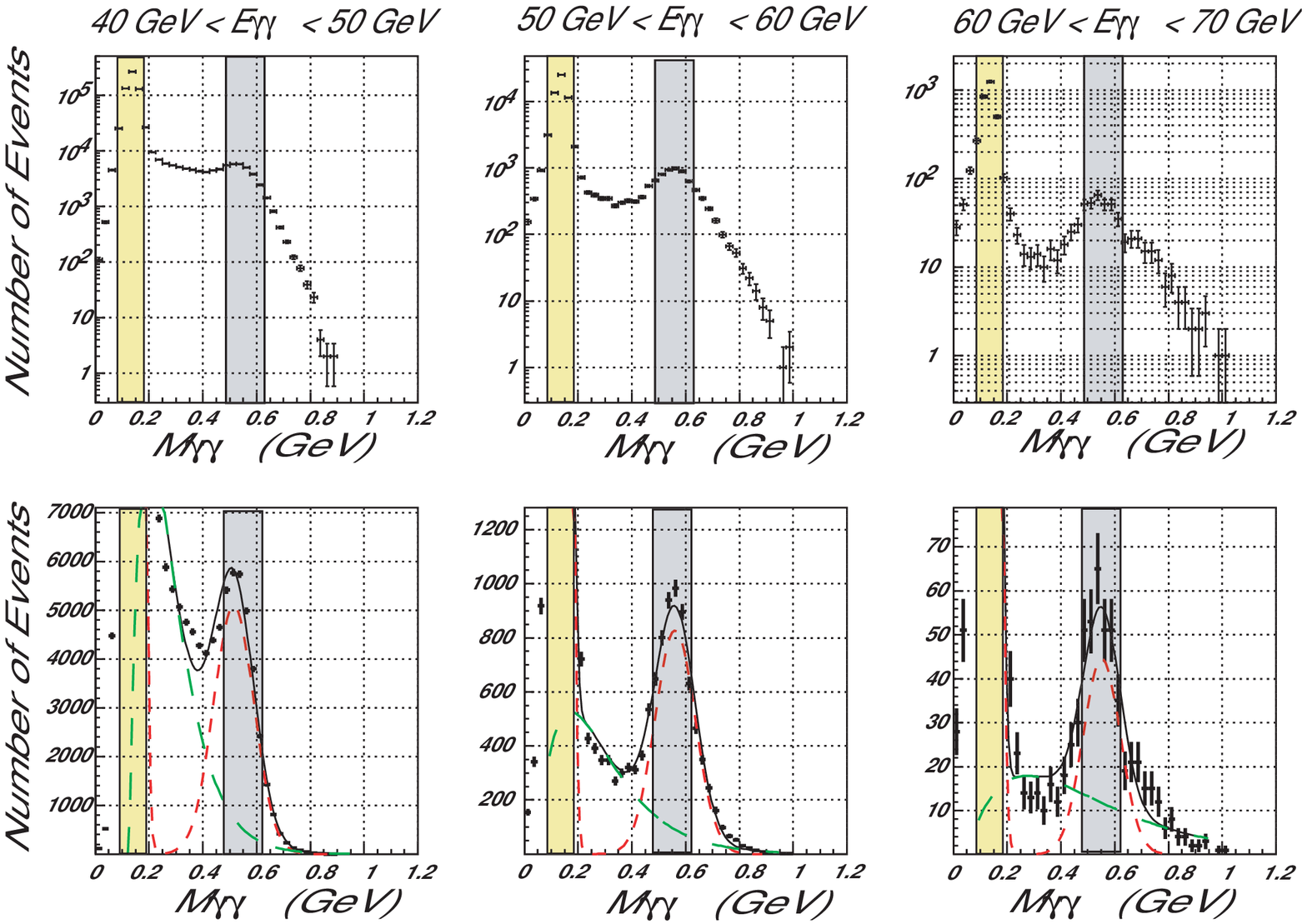}
\caption{Two photon mass distributions are shown in 3 energy ranges
with
$\pi^0$ and $\eta$ mass bands indicated.  
Note the log scale on the upper plots and that the 
linear scale on the lower plots 
emphasizes the $\eta$ peak.}
\label{fig:massdist}
\end{minipage}
\hspace{.15in}
\begin{minipage}[b]{2.1in}
\includegraphics*[width=2.1in]{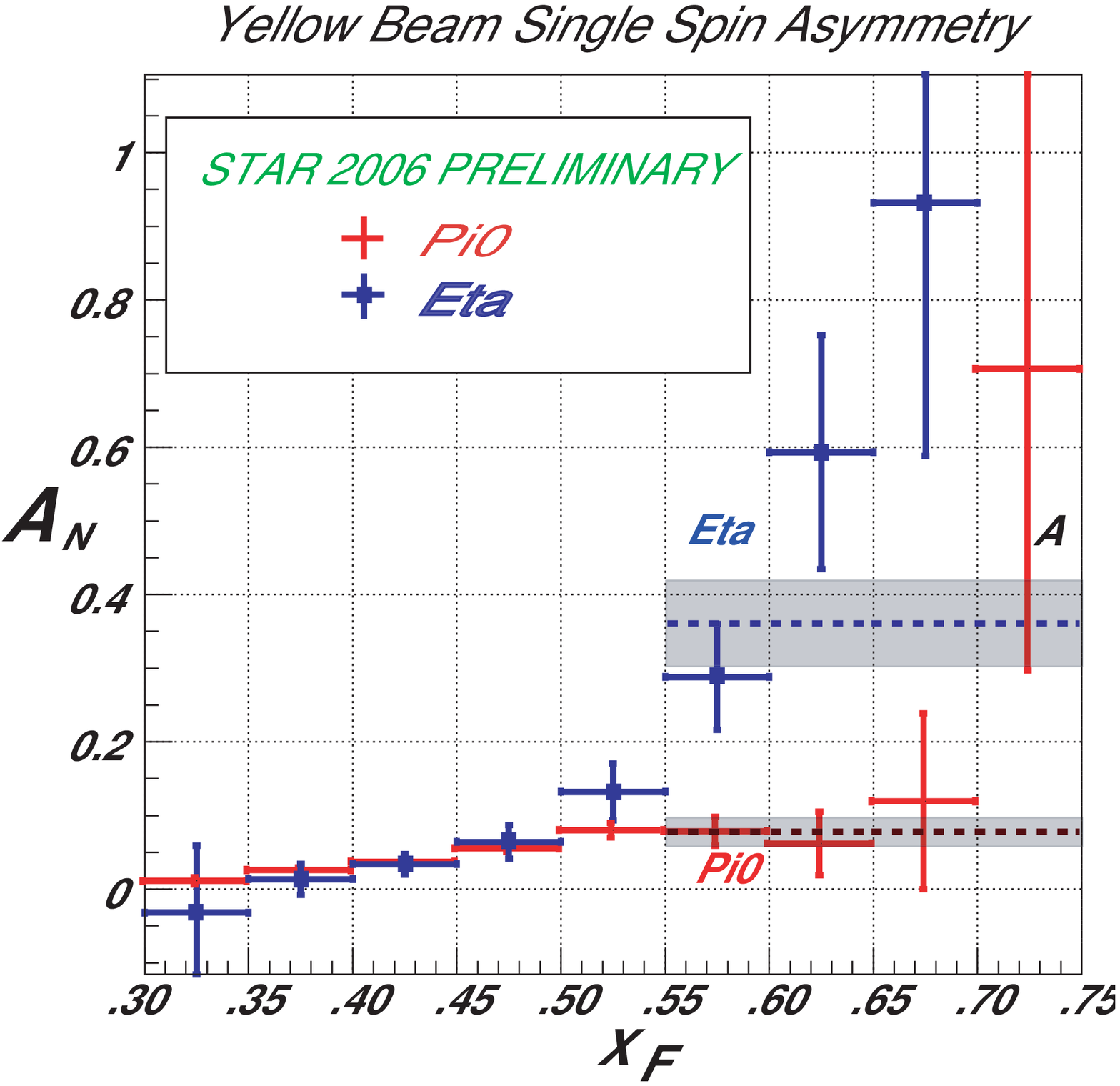}
\caption{The dependence of $A_N$ on $x_F$ is plotted
for the $\pi^0$ and $\eta$ mass bands
defined by $m_{\pi^0}=0.135\pm 0.05$ GeV for $\pi^0$ and 
$m_{\eta}=0.55\pm 0.07$ GeV for $\eta$.}
\label{fig:andist} 
\end{minipage}
\end{figure}

With the STAR Forward Pion Detector (FPD), during the 2006 RHIC run with 
transversely polarized protons of energy $\sqrt{s}=200\ GeV$, the transverse 
single spin asymmetry for $\pi^0$ 
production at large Feynman $x_F$ has already been 
reported\cite{:2008qb}.
Using the same FPD apparatus and same data set, the corresponding $A_N$ 
asymmetry is now shown for photon pairs in the $\eta$ meson region. The     
selection of events is chosen to optimize FPD acceptance and yield 
for $\eta$ mesons relative to $\pi^0$s, cutting on two photon
pseudo-rapidity (Y) and 
azimuthal angle ($\phi$), 
$((Y-3.65)^2+tan(\phi)^2)<(0.15)^2$. 
This cut corresponds to the centers of the two FPD calorimeters located 
downstream, both left and right, of the yellow polarized RHIC beam. 
The transverse polarization axis 
is oriented at $\phi=\pm\frac{\pi}{2}$ with an average
polarization for the yellow beam of $56\%$. The FPD acceptance for 
two photon decays  improves for larger $x_F$ in the 
$\eta$ mass region and is significant for $x_F$ above about
0.4. In Figure \ref{fig:massdist}, the three pairs of mass plots 
correspond to a further selection of events with photon pair energies 
in the indicated ranges. The mass regions that will be associated with 
$\pi^0$ and $\eta$ mesons 
are indicated with the vertical bands. 
It is for the events in these two mass bands that
the single spin asymmetry is calculated as a function of $x_F$ 
using the cross ratio 
method\cite{:2008qb}. These asymmetries are shown in Figure \ref{fig:andist}.

\section{Results and Summary}

Like the E704 measurement, the asymmetry $A_N$ for the $\eta$ meson mass region
is larger than that for the $\pi^0$ region. The weighted average
of this asymmetry over
Feynman $x_F$ in the range $0.55<x_F<.75$ is 
$\langle A_N\rangle _{\eta}=0.361\pm 0.064$. 
In comparison, for the $\pi^0$ mass region
the corresponding asymmetry is 
$\langle A_N\rangle_{\pi^0}=0.078\pm 0.018$. 
The errors here are statistical, with preliminary estimates of
systematic errors much smaller.
This difference
in $A_N$ between the $\eta$ mass region and the $\pi^0$ region is more than 
four standard deviations.
In the $\eta$ region, we further note the trend to 
higher asymmetry at
larger $x_F$, raising the question as to whether the asymmetry is 
approaching a maximal value of 1 in the high $x_F$ end of this range.
This result is consistent with the E704 measurement but is more significant 
with largest $A_N$ at higher $x_F$ than previously measured.

%
%
%
\bibliographystyle{unsrt}
\bibliography{heppel}

\begin{thebibliography}{1}

\bibitem{Adams:2006uz}
J.~Adams et~al.
\newblock {\em Phys. Rev. Lett.}, 97:152302, 2006.

\bibitem{Sivers:1990fh}
D.~W. Sivers.
\newblock {\em Phys. Rev.}, D43:261--263, 1991.

\bibitem{D'Alesio:2004up}
U.~D'Alesio and F.~Murgia.
\newblock {\em Phys. Rev.}, D70:074009, 2004.

\bibitem{Collins:1993kq}
J.~C. Collins, S.~F. Heppelmann, and G.~A. Ladinsky.
\newblock {\em Nucl. Phys.}, B420:565--582, 1994.

\bibitem{Bramon:1997va}
A.~Bramon, R.~Escribano, and M.~D. Scadron.
\newblock {\em Eur. Phys. J.}, C7:271--278, 1999.

\bibitem{Adams:1997dp}
D.~L.~Adams et~al.
\newblock {\em Nucl. Phys.}, B510:3--11, 1998.

\bibitem{:2008qb}
B.~I.~Abelev et~al.
\newblock {\em Phys. Rev. Lett.}, 101:222001, 2008.

\end{thebibliography}

%
%
%
%
%
%
\end{document}